# Distribution function of the random field and polar properties of the relaxor ferroelectric films.

E.A.Eliseev*, M.D.Glinchuk

*Institute for Problems of Materials Science, National Academy of Science of Ukraine, Krjijanovskogo 3, 03142 Kiev, Ukraine,*
*eliseev@mail.i.com.ua

The model for calculation of relaxor ferroelectrics thin films properties is proposed. The basis of the model is the theory of random field. This field is originated from the chemical disorder allowing for influence of the film surfaces, which destroys the polar long-range order and transform it into mixed state of ferroelectric glass (FG) or dipole glass (DG). The spatial profiles and averaged over coordinate inside the film values of properties of relaxor ferroelectric film were calculated with the random field distribution function. As an example the dependence of the order parameter on film thickness, temperature and distribution function characteristics was obtained. The critical thickness and temperature of the size-driven phase transition from FG to DG state as a function of the film and distribution function characteristics were calculated.

## 1. INTRODUCTION

Relaxor ferroelectric thin films attract much attention of scientists in the last years (see e.g. [1-3]). Bulk relaxor ferroelectrics possess excellent dielectric, electromechanical and electrooptical properties which found numerical technical applications. The experimental study of relaxor ferroelectric thin films are aimed to find out if it is possible to obtain the same or even improved properties by the changing film thickness, substrate material and growth conditions. Because of the complexity of the problem there are no answers about the influence of the aforementioned external and internal factors, e.g. chemical disorder, on the relaxor ferroelectric films properties. It was found that these properties strongly differ from those of bulk relaxor ferroelectrics and common ferroelectric films [1].

Situation appeared more complex because in the majority of works the polycrystalline films with nanosize grains were studied (see, e.g. [1, 2, 4]). In one of the last publication devoted to experimental study of epitaxial relaxor ferroelectric film the influence of the external electric field on the dynamic dielectric properties was studied [5]. The obtained information about low energy barriers between stable states of the reorientable polar clusters in the film compared to those in the bulk material shed no light on the static properties of the film.

Aforementioned factors made it cumbersome to find out the real features of pure relaxor ferroelectrics thin films, which can be more clearly studied in epitaxial films. Maybe because of the problem complexity and restricted experimental data there no theoretical works with description of theses film properties behaviour. It is obvious that the development of the theory will be the great help for experimenters and engineers in their efforts to understand correctly the expected properties and the way of their observations.

In the present paper we proposed for the first time random field based model of relaxor ferroelectrics thin films. Similarly to the model proposed earlier for the bulk relaxor ferroelectrics [6] the ferroelectric long range order of the system is destroyed by the random filed originated from the chemical disorder, in the thin films it is influenced by the surface effects. The calculation of the random field distribution function is performed in the statistical theory framework [7-9]. Then one can calculate the properties of the relaxor ferroelectrics by averaging the observable quantities of the ordered system on the electric field with random field distribution functions. For the sake of illustration we calculated the spatial profile of the order parameter, its average value dependence on the film thickness, temperature and distribution function width as well as critical thickness of the size-driven phase transition from relaxor to dipole glass state. Hereafter we consider epitaxial monodomain film of insulating ferroelectric material with ideal superconducting electrodes.

## 2. RANDOM FIELD DISTRIBUTION FUNCTION

This function can be calculated in the statistical the statistical theory framework similarly to Ref. [10, 11]. However in the thin film the contribution of the surfaces should be taken into account. In what follows we consider randomly distributed charges and electric dipoles as the sources of random electric field. The surface influence on the distribution function for these sources will be considered by taking into account the image charges and dipoles [12] similarly to our previous work [13].

Thus the distribution function of the random field can be introduced by the following way [10]:

$$f(\vec{E}) = \overline{\left\langle\left\langle \delta\!\left(\vec{E} - \vec{E}(\vec{r}_i)\right)\right\rangle\right\rangle} \qquad (1)$$

where $\vec{E}(\vec{r}_i)$ is local electric random field in the point $\vec{r}_i$, which in general case includes the contributions both the real random field sources and its images. The bar denotes averaging over spatial configurations of random field sources, $\langle\langle \ldots \rangle\rangle$ means thermal averaging over orientations of random dipoles so that the distribution function is expressed through itself in a self-consistent manner. The calculation of $f(\vec{E})$ on the basis of expression (1) was carried out in [10] with the help of the statistical method [13]. It yields

$$f(\vec{E}) = \frac{1}{(2\pi)^3} \iiint\limits_{-\infty}^{+\infty} \exp\!\left(i\vec{\rho}\cdot\vec{E} - \sum_{m=1}^{k} F_m(\vec{\rho})\right) d^3\rho \qquad (2)$$

$$F_m(\vec{\rho}) = n_m \int\limits_V \left\langle\left\langle \exp\!\left(-i\vec{\rho}\cdot\vec{E}_m(\vec{r})\right) - 1\right\rangle\right\rangle d^3r \qquad (3)$$

where $\vec{E}_m(\vec{r})$ is an electric field in the point $\vec{r}$, created by *m*-th type of random field sources with concentration $n_m$, all the sources from $m = 1$ to $m = k$ are supposed to be independent.

Hereafter we consider the system of dipoles which can be directed either along or against one preferential axis. That is why we can suggest the random electric field oriented along this axis because the perpendicular to dipole components can not influence their behavior. We choose z-axis as the preferential one and vector $\vec{\rho}$ as follows $(0,0,\rho)$.

It is easy to show that for the high enough concentration of random field sources the Gaussian approximation can be used for the calculation of the distribution function

characteristics. Namely, the real and imaginary parts of expression (3) can be represented in the form:

$$\text{Re}(F_m(\rho)) \approx \rho^2 \frac{n_m}{2} \int_V E_{mz}(\vec{r})^2 dV \equiv \Delta_m \rho^2 \quad (4a)$$

$$\text{Im}(F_m(\rho)) \approx \rho n_m \int_V E_{mz}(\vec{r}) dV \equiv E_{0m} \rho \quad (4b)$$

Quantity $\Delta_m$ determines the contribution of the source of type $m$ to the widths of distribution function. The width $\Delta E$ of the random electric field distribution function can be written in the following form:

$$\Delta E = \sqrt{\sum_m \Delta_m} . \quad (5)$$

Hereinafter we consider random field two type of sources such as defect charges and dipoles. It was shown recently [13] that the mean field value is independent on the film thickness and coincides with the one for the bulk system:

$$E_0 = 4\pi n_2 d_z^* / \varepsilon_i \quad (6)$$

## 4. CHARGED DEFETS CONTRIBUTION

Let us consider the dependence of distribution function width on the coordinate inside the film of relaxor ferroelectrics. It has thickness $h$, permittivity $\varepsilon_i$ and placed between electrodes with fixed potentials. We choose Z-axis normally to the film surfaces and film region is $z \in [0, h]$. It is known that the electric field of charge q situated inside the film between two electrodes can be written as the sum of field of point charge q in the infinite medium with dielectric permittivity $\varepsilon_i$ and contributions of image charges [12]:

$$\vec{E}_1 = \frac{q}{\varepsilon_i} \begin{cases} \dfrac{\vec{r}_0}{r_0^3} - \dfrac{\vec{r}_{1+}}{r_{1+}^3} - \dfrac{\vec{r}_{1-}}{r_{1-}^3} + ..., & 0 \leq z \leq h; \\ 0, & \{z<0\} \cup \{z>h\}; \end{cases} \quad (7)$$

Here $\vec{r}_0, \vec{r}_{1+}, \vec{r}_{1-}$ are the radius vectors from the points, where the real charge and its first two images are situated respectively. The structure of the omitted terms is obvious.

In order to consider the electric field (7) in the Gaussian approximation we have to suppose that charged sources have finite sizes. For the spherical form defects with radius $b$ and the full charge $q$ homogeneously distributed on the surface the electric field outside the sphere can be represented by (7), while inside the sphere in the expression (7) the first term should be omitted.

The calculations of the integral (4b) with Eq. (7) had shown that $E_{01}=0$, so that the defect charges in the relaxor films do not contribute to the most probable electric field as well as in infinite (bulk) system [10].

The quantity $\Delta_1$ determining the contribution of the defect charges to the distribution function width can be found only as a series on image charges:

$$\Delta_1 = \frac{n_1}{2} \sum_{m,l} (2-\delta_{ml}) \frac{q_m q_l}{\varepsilon_i^2} \iiint_{\substack{z \in [-h/2, h/2] \\ |\vec{r}-\vec{r}_0| \geq b}} \frac{(z-z_m)(z-z_l)}{|\vec{r}-\vec{r}_m|^3 |\vec{r}-\vec{r}_l|^3} dV \qquad (8)$$

Here summation is performed over all the image charges and one real charge, $\delta_{ml}$ is the Kroneker delta, $\vec{r}_{m,l}$ and $q_{m,l}$ are the coordinates and effective charges of *m*-th and *l*-th charge respectively. The coordinate of real charge is $\vec{r}_0 = (0, 0, z_0)$. After the simple but cumbersome calculation the integrals from (8) can expressed in terms of the rational functions of charge coordinates.

When using (8) we cut off the series after the nearest eight image charges. The contribution of other terms is no more than several percents.

For the charges far from surfaces in the film with the infinitely large thickness $h \to \infty$ one can obtain

$$\Delta_1^\infty = \frac{2\pi}{3} \frac{n_1}{b} \left(\frac{q}{\varepsilon_i}\right)^2. \qquad (9)$$

It is seen that in the bulk system the distribution function is isotropic for the charged defects as random field sources.

Recently [13] we have proposed an approximate approach for the calculation of the random field distribution function in the relaxor ferroelectrics films with the large enough thickness (*h*>>*b*). We take into account an image charges from the bottom and upper surface separately. Allowing for the electric field decrease with distance we do not take into account the other image and developed the following approximate formula:

$$\Delta_{appr}(z_0, h) \cong \frac{\Delta_{1S}(h-z_0)\Delta_{1S}(z_0)}{\Delta_1^\infty}, \qquad \Delta_{1S}(a) = \Delta_1^\infty \left[1 - \frac{b^3}{8a^3}\right]. \qquad (10)$$

Here $\Delta_S(a)$ determine the width of the random field distribution function for the charged defect at the distance *a* from the flat boundary between dielectric with permittivity $\varepsilon_i$ and metallic electrode [13].

The comparison between the exact expression (8) and approximate one (10) had shown that they practically coincide with each other. That is why we will use hereinafter only the approximate expression (10).

## 5. ELECTRIC DIPOLES

In this section we consider randomly distributed and oriented electric dipoles as a random field sources in highly polarizable dielectric with soft polar mode. It was shown earlier [10] that for the dipoles with the arm much smaller than lattice constant indirect interaction via soft mode phonons of host lattice between embedded dipoles leads to the renormalization of the dipole electric field.

As mentioned before, for source inside the film the electric field is the field caused by the real source and its images [12]. We have shown in the previous section that for the thick enough films one could neglect the influence of one surface near the other one. The electric fields caused by dipole sources decreases with distance even faster than the one created by the charges. Therefore we can find the contribution to the random field distribution function halfwidth from the dipoles in the vicinity of the film surfaces and then use the approximate expression similar to Eq. (10) for the contribution of the dipoles inside the film.

So, let us consider a dipole situated near the boundary between two media. The coordinate origin coincides with the dipole location. The half-space $z \geq a > 0$ is metallic, and the half-space $z < a$ is filled with the highly polarizable dielectric with permittivity $\varepsilon_i$. For this system one has to take into account the depolarization field arising due to the free charges localized on the film surfaces. This field is proportional to the system polarization components normal to its surface. Taking into account that the electric field is created by a real source and its images one can write it as follows [10]:

$$\vec{E} = \frac{1}{\varepsilon_i} \begin{cases} \hat{K}(\vec{r})\vec{d}^* - \hat{K}(\vec{r} - 2a\vec{e}_z)\vec{d}' + \vec{e}_z 4\pi \frac{d_z^*}{V}, & z < a; \\ 0 & z \geq a; \end{cases} \qquad (11)$$

$$K_{nm}(\vec{r}) = \delta_{nm} \frac{\exp(-r/r_c)}{r_c^2 r} + \frac{\partial^2}{\partial n \partial m}\left(\frac{1 - \exp(-r/r_c)}{r}\right). \qquad (12)$$

Here the components of the real dipole $\vec{d}^*$ and its image $\vec{d}'$ moments are $(d_x^*, d_y^*, d_z^*)$ and $(d_x^*, d_y^*, -d_z^*)$ respectively, $\vec{e}_z$ is the unit vector of z-axis, V is the volume of the sample, the last term from Eq.(11) is the contribution of depolarization field, $r_c$ is the correlation radius of the host lattice polar phonons fluctuations.

Hereafter we calculate the quantity (4a) determining the width of the distribution function for the electric dipoles perpendicular to the boundary plane. The calculations of $\Delta_2$ with the field distribution (11) yields:

$$\Delta_{2S}(a) = \frac{n_2}{r_c^3}\left(\frac{d_z^*}{\varepsilon_i}\right)^2 \left(\frac{8\pi}{15} + r_c^3 \iiint_{z \leq a} dV\, K_{zz}(\vec{r}) K_{zz}(\vec{r} - 2a\vec{e}_z)\right) \qquad (13)$$

The last integral depends only on the dimensionless distance from the surface $a/r_c$. It was calculated numerically. At the distance $a \gg r_c$ (at least in the case $a > 3r_c$) it can be easily expressed in terms of elementary functions:

$$\Delta_{2S}(a \gg r_c) \cong \Delta_2^\infty \left(1 + \frac{15}{8}\left(\frac{r_c}{a}\right)^3\left[1 - 3\left(\frac{r_c}{a}\right)^2\right]\right), \quad a \gg r_c. \qquad (14)$$

$$\Delta_2^\infty = \frac{8\pi}{15}\frac{n_2}{r_c^3}\left(\frac{d_z^*}{\varepsilon_i}\right)^2 \qquad (15)$$

In order to obtain the distribution of $\Delta_2$ inside the film we use the approximate expression similar to Eq. (10).
The width of the random field distribution function $\Delta E$ consists of the two terms: $\Delta E = \sqrt{\Delta_1 + \Delta_2}$. Here summands $\Delta_{1,2}$ correspond to the contributions of the defect charges (10) and defect dipoles (13). They depend on the distance from the film surfaces because of the image charges influence, their magnitudes are determined by the bulk values $\Delta_1^\infty$ and $\Delta_2^\infty$ from (9) and (15). Hereinafter we use the following designations:

$$\Delta E_1 = \sqrt{\Delta_1^\infty}, \quad \Delta E_2 = \sqrt{\Delta_2^\infty}, \quad \Delta E_\infty = \sqrt{\Delta_1^\infty + \Delta_2^\infty}, \tag{16}$$

The dependence of $\Delta E$ on the coordinate $z$ is represented in Fig. 1 for the two cases. The contribution of charges or dipole sources is prevailing for Fig. 1a or Fig. 1b respectively. It is seen that $\Delta E$ near the surface of the film can be either smaller or higher than its bulk value. One can see that the distribution function width tends to the bulk width value in the central part of a film with the film thickness increase.

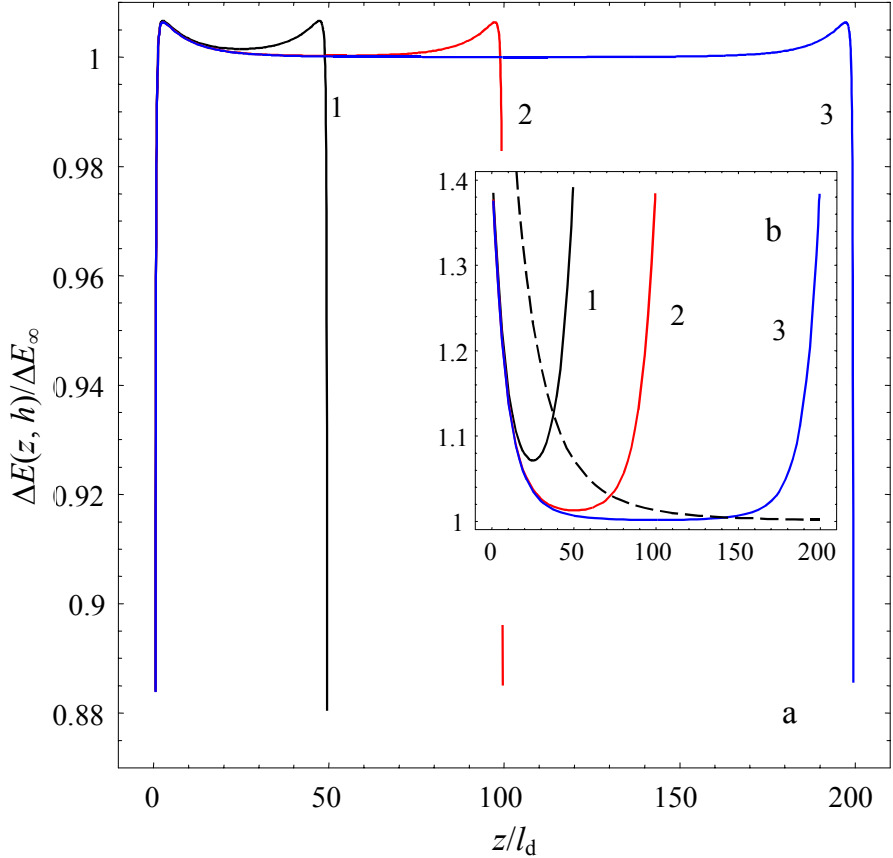

**Figure 1**. Distribution function width dependence on the coordinate inside the film $r_c/l_d=10$, $b/l_d=0.5$, $h/l_d=50$, 100, 200 (solid curves 1, 2, 3), $\Delta E_1/\Delta E_2=7$ (a), 0.14 (b). Dashed curve represent dependence of $\Delta E$ in the film center ($z=h/2$) on the film thickness

## 3. RELAXOR FERROELECTRIC FILM PROPERTIES.

The obtained distribution function permits to calculate all the properties of relaxor ferroelectric film. The observable physical quantity can be found as the average with random field distribution functions $f(\vec{E})$ of the quantity of the ordered system calculated for given external electric field, namely:

$$\langle A(z,T)\rangle = \int_{-\infty}^{+\infty} f(\vec{E}) A(\vec{E},z,T) d\vec{E} \tag{17}$$

Here $A(\vec{E},z,T)$ is the observable physical quantity of the ordered system, dependent on the electric field, coordinate and temperature; $\vec{E}$ is the random electric field. As an example we will consider the static order parameter, which determines the polarization of the system.

The ordered system of dipole with two possible orientations can be described by the order parameter $L$ as the ratio of the coherently oriented dipoles. This system represents ferroelectrics of order-disorder type. The films of this type material were considered earlier

[14] for the dipoles perpendicular to the film surfaces. It is shown, that order parameter $L$ for this system can be found as

$$L = \tanh\left(\frac{d^*E_z}{k_BT} + \frac{d^*E_0}{k_BT}\left(1-\frac{h_0}{h}\right)L\right), \quad h_0 = \frac{2\delta}{\lambda+l_d}, \quad l_d \approx \sqrt{E_0\delta/4\pi n(d^*)^2} \quad (18a)$$

Here $d^*$ is the effective dipole moment, $E_z$ and $E_0$ are z-components of the electric field and mean field respectively, z axis is perpendicular to the film surface, $h$ is the film thickness, $h_0$ is critical value of the thickness, $\delta$ determines the correlation energy, $\lambda$ is the extrapolation length, $n$ is the concentration of dipoles [14]. The quantity $4\pi n(d^*)^2/k_B$ is equal to the Curie – Weiss constant $C_{CW}$ of the ordered material.

The distribution of order parameter across the film is follows [14]:

$$l(z) = L(1-\varphi(z)), \quad \varphi(z) = \frac{\cosh(z/l_d)}{\cosh(h/2l_d)+\sinh(h/2l_d)\lambda/l_d}. \quad (18b)$$

For the films of relaxor ferroelectrics one has to average order parameter (18) on the random electric field created by different sources [10]. Taking into account Eq. (17), the distribution function (2) and expressions (18) we can write the following equation for the order parameter $L$:

$$L = \int_{-\infty}^{+\infty} f(\vec{E})\tanh\left(\frac{d^*E_z}{k_BT} + \frac{d^*E_0}{k_BT}\left(1-\frac{h_0}{h}\right)L\right)d\vec{E} \quad (20)$$

Using the distribution function width dependence on the coordinate inside the film, one can obtain with the help of Eqs. (18), (19) the distribution of order parameter inside the film (Fig. 2, 3).

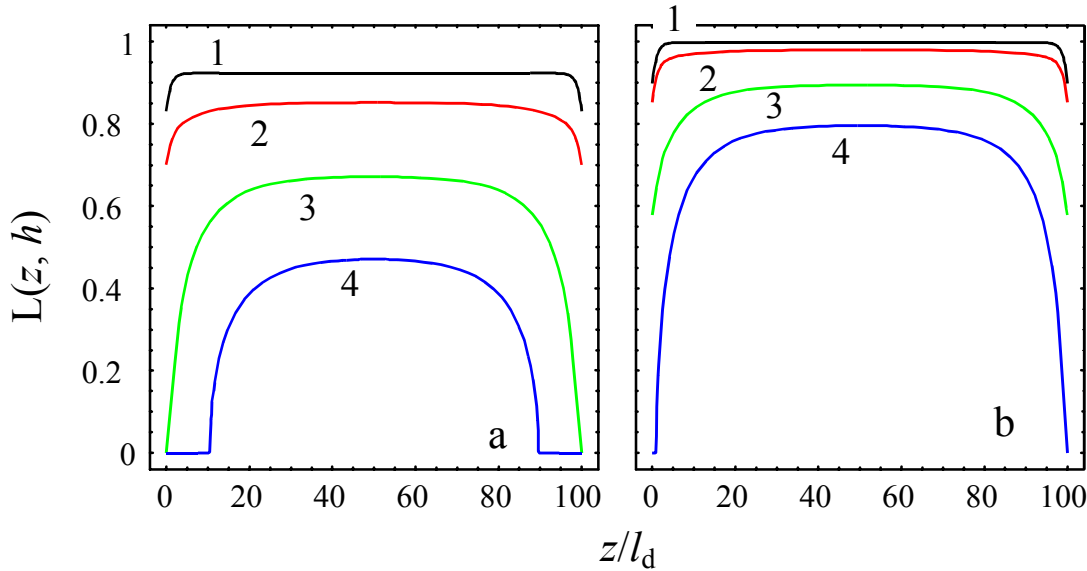

**Figure 2**. Order parameter distribution inside the film, $r_c/l_d=10$, $b/l_d=0.5$, $\lambda/l_d=9$, $h/l_d=100$, $C_{CW}/T_d=100$, $T/T_d=0.5$ (a), 0.3 (b). $\Delta E=0$ (curve 1), $\Delta E_1/E_{0\infty}=0.1$, $\Delta E_2/E_{0\infty}=0.3, 0.5, 0.6$ (curves 2, 3, 4).

The temperature is normalized on the so called Burns temperature $T_d$ [6, 9] which coincides with transition temperature of the bulk ordered system $d^*E_0/k_B$ in the framework of this model.

It is seen from Fig. 2 that at fixed film thickness and temperature order parameter essentially depends on the width of the distribution function $\Delta E$. With its amplitude $\Delta E_\infty$ increases the order parameter decreases. However the behavior at the film surfaces depends on the ratio of the contributions of different sources $\Delta E_1$ and $\Delta E_2$. When the contribution of charge sources is small, i.e. $\Delta E_1 \ll \Delta E_2$ (see Fig. 2), the order parameter in these regions decreases faster than in the film center. Moreover, for large enough $\Delta E_2$ there is a layer near the surface with L=0, while in the film center L≠0 (see curve 4 in Fig. 2a).

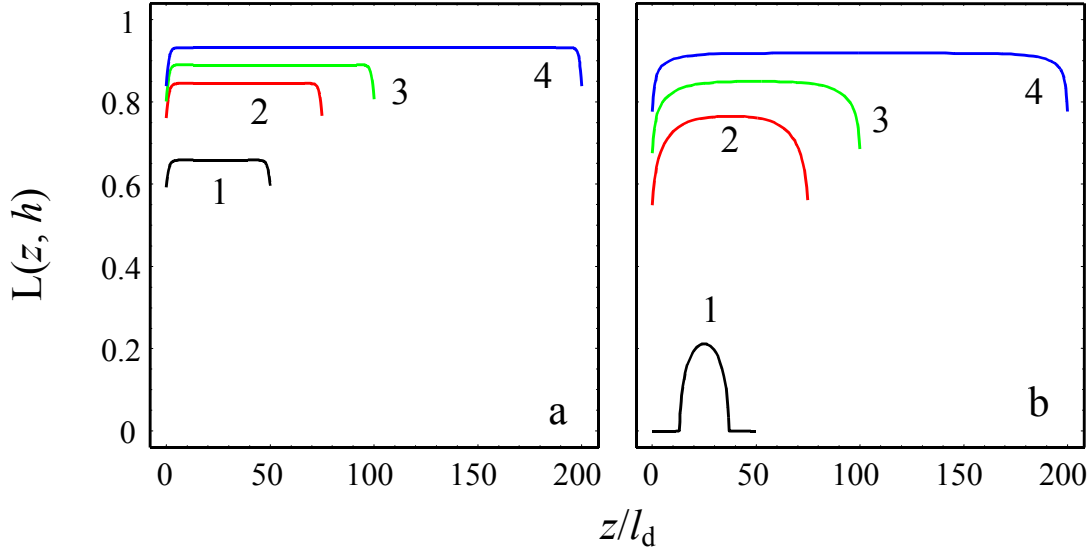

**Figure 3**. Order parameter distribution inside the film, $r_c/l_d=10$, $b/l_d=0.5$, $\lambda/l_d=9$, $h/l_d=50, 75, 100, 200$ (curves 1, 2, 3, 4), $C_{CW}/T_d=100$, $T/T_d=0.5$, $\Delta E=0$ (a), $\Delta E_2/E_{0\infty}=0.32$, $\Delta E_1/E_{0\infty}=0$ (b).

When the thickness of the film decreases and the temperature is kept fixed, the size-driven phase transition is known to take place in ordered ferroelectrics (see e.g. [14, 15]). It is clearly seen from Fig. 3a that the order parameter amplitude decreases with the film thickness decrease. For the films of disordered ferroelectrics situation is more complex. The order parameter $L$ decreases faster near the film surfaces (see Fig. 3b).

For the sake of comparison of the order parameter in the relaxor ferroelectric films and in bulk material we calculated the averaged over coordinate $z$ the order parameter $\overline{L}$. Its dependences on the film thickness and temperature are represented in Figs. 4 and 5 respectively for the different values of the distribution width $\Delta E$ which determines the degree of disorder of relaxor ferroelectrics, i.e. it increases with $\Delta E$ increase [6].

It is seen that for the bulk material $\overline{L}$ (see circles in Fig. 4 and dashed line in Fig. 5) is larger than for the films with all other parameters fixed. For the given thickness of the film $\overline{L}$ value decreases with degree of disorder increase (see crosses in Fig. 4), the largest $\overline{L}$ value being characteristic for the completely ordered film (curve 1 in Fig. 4).

The size driven phase transition from the mixed state ferroelectric glass (ferroglass, FG; $\overline{L} \neq 0$, $\overline{L^2} \neq 0$) to the dipole glass state (DG, $\overline{L} = 0$, $\overline{L^2} \neq 0$) is clearly seen from Figs. 4 and 5 as the critical values of temperature $T_{cr}$ and thickness $h_{cr}$ respectively where order parameter $\overline{L}$ tends to zero. The dependence of critical parameters on the degree of disorder is obvious, namely:

$$T_{cr}^{Ordered\ Films} > T_{cr}^{Bulk\ relaxor} > T_{cr}^{Relaxor\ Films} \qquad (20)$$

$$h_{cr}^{Ordered\ Films} < h_{cr}^{Bulk\ relaxor} < h_{cr}^{Relaxor\ Films} \qquad (21)$$

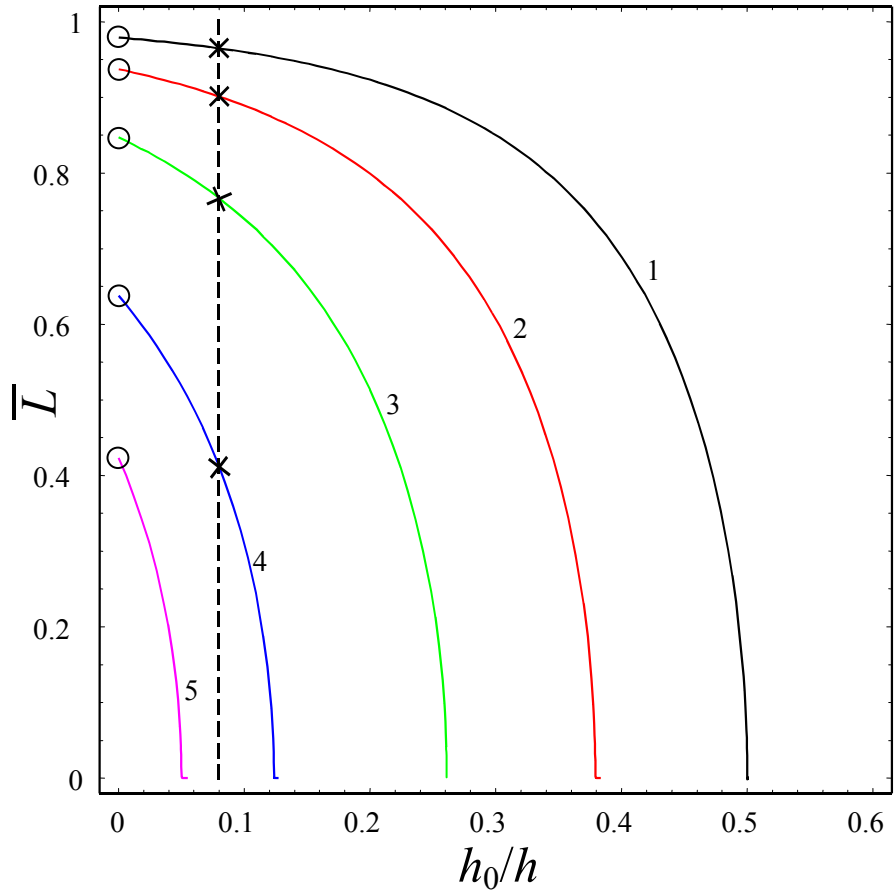

**Figure 4**. Dependence of the mean value of the order parameter on the reciprocal thickness temperature for the following parameters $T/T_d$=0.5, $\Delta E/E_{0\infty}$=0, 0.4, 0.6, 0.8, 0.9 (curves 1, 2, 3, 4, 5).

Qualitatively one can expect the obtained dependence of $\overline{L}$, $h_{cr}$ and $T_{cr}$ values on the degree of disorder because the increase of the disorder in the relaxor films in comparison to ordered films and to bulk relaxor ferroelectrics must destroy polar long range more effectively, leading to DG state for thicker film and for smaller temperature. The curves in Fig. 4, 5 represent this behavior qualitatively.

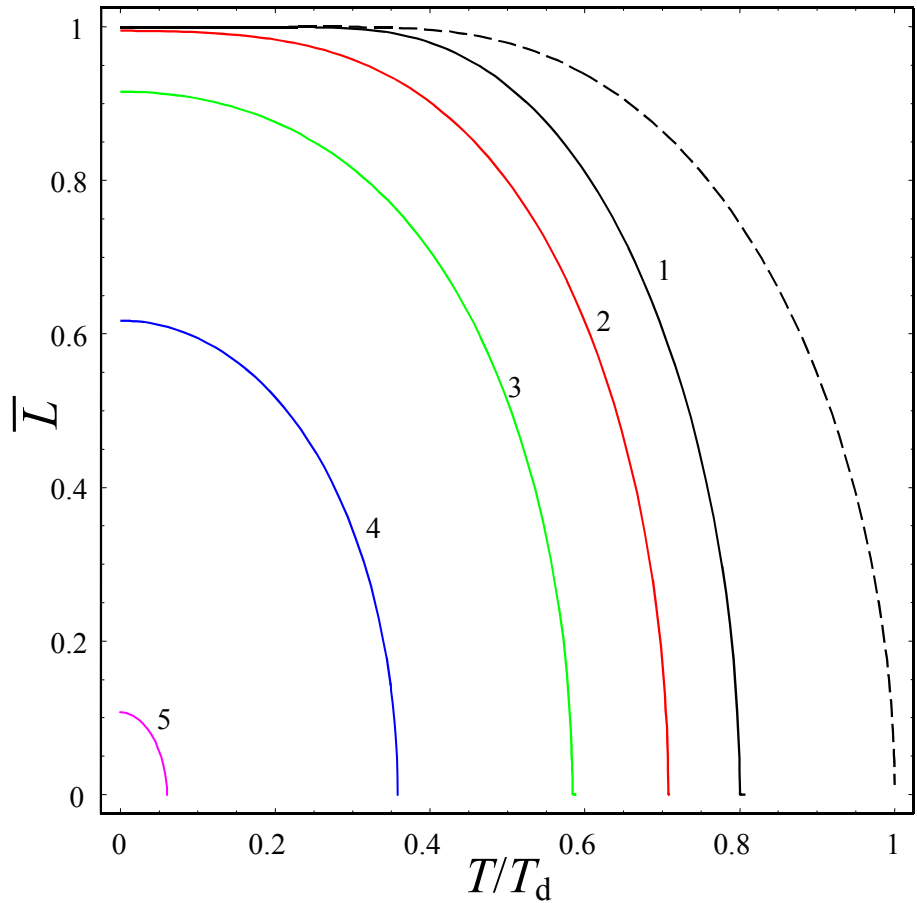

**Figure 5**. Dependence of the mean value of the order parameter on temperature for the following parameters $h/h_0=5$, $\Delta E/E_{0\infty}=0$, 0.4, 0.6, 0.8, 0.9 (solid curves 1, 2, 3, 4, 5). Dashed curve represents the order parameter for the bulk ordered ferroelectrics.

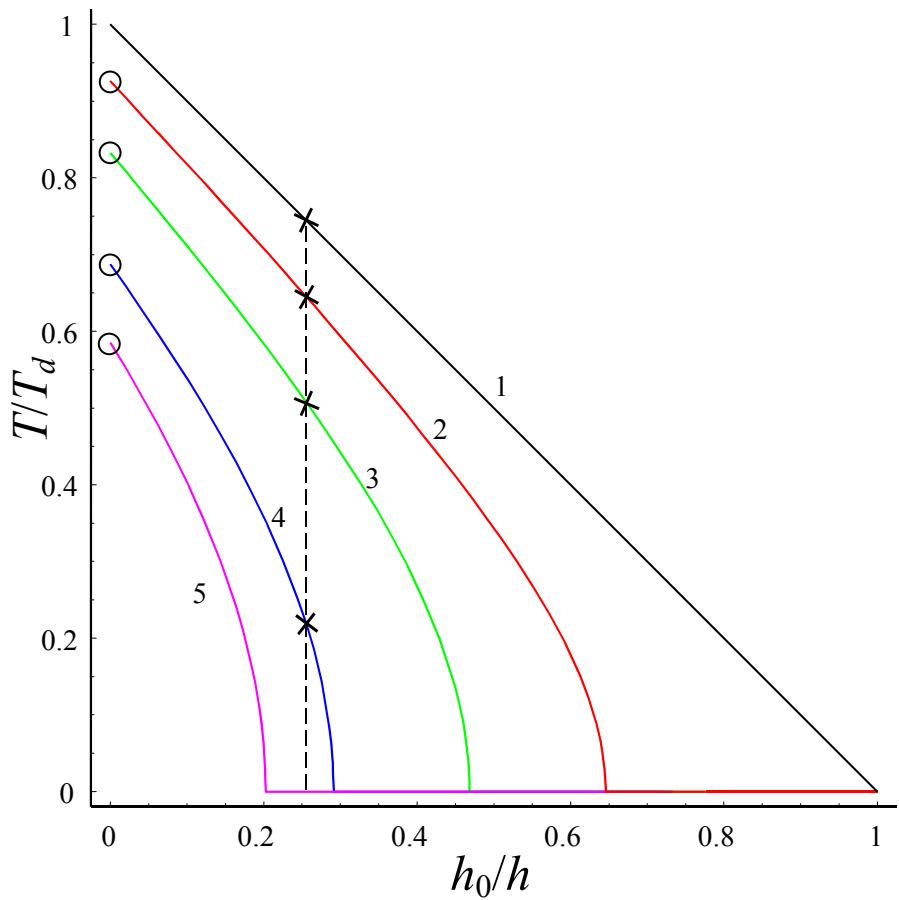

**Figure 6**. The phase diagram temperature - film thickness for the following parameters $\Delta E/E_{0\infty}=0$, 0.4, 0.6, 0.8, 0.9 (curves 1, 2, 3, 4, 5).

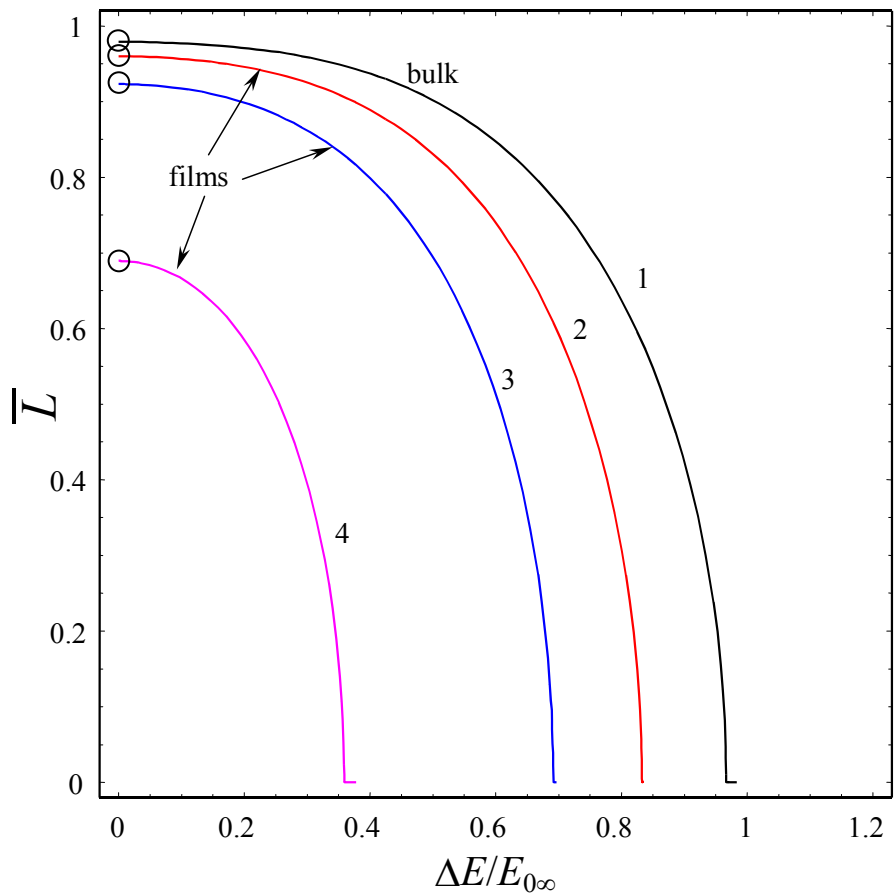

**Figure 7**. Averaged order parameter dependence on the degree of disorder for the following parameters: $h/h_0=\infty$, 10, 5, 2.5 (curves 1, 2, 3, 4), $T/T_d=0.5$.

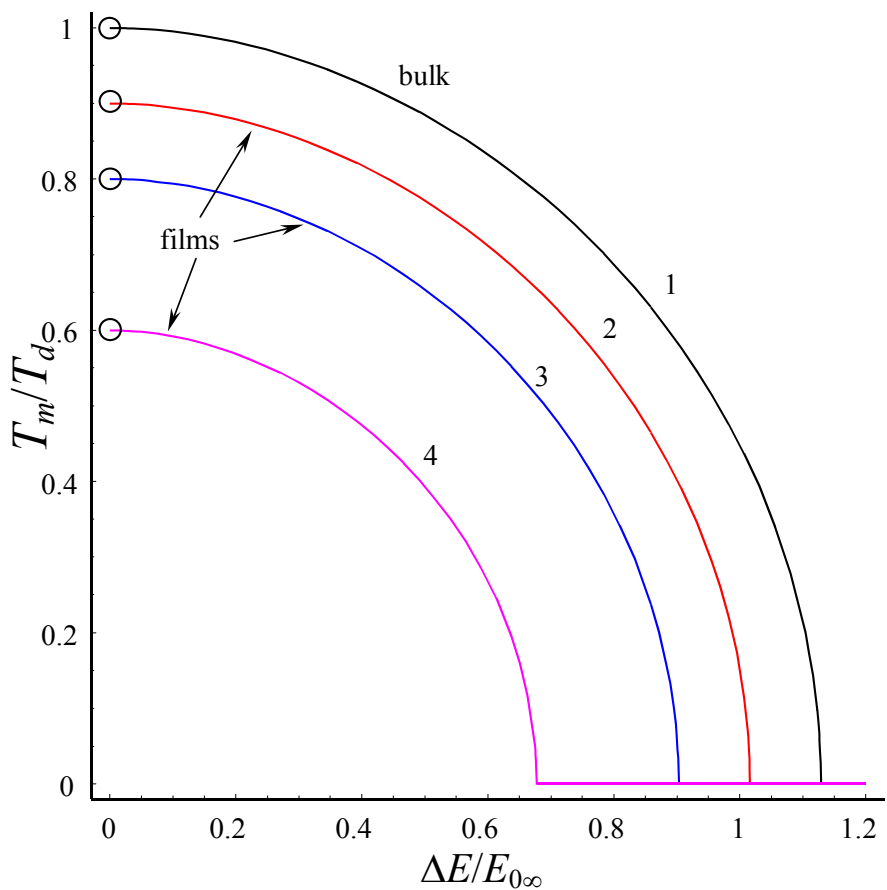

**Figure 8**. Temperature of susceptibility maximum dependence on the degree of disorder for the following parameters: $h/h_0=\infty$, 10, 5, 2.5 (curves 1, 2, 3, 4).

The same type of behavior can be seen from Fig. 6 where the phase diagram of the system is depicted. This diagram represents the dependence of the temperature of the dielectric susceptibility maximum $T_m$ on reciprocal thickness of the film. The decrease of $T_m$ with the increase of disorder (width $\Delta E$) is evident, the value $T_m$ in the relaxor film being smaller than in the ordered film and than in the bulk relaxor (circles). The dependence of $\overline{L}$ on the degree of disorder is depicted in Fig. 7. It is seen that at fixed temperature and film thickness there is a critical value of $\Delta E$ when $\overline{L}$ tends to zero. Another type of phase diagrams is represented in Fig. 8 as the dependence of the temperature $T_m$ on the distribution function width $\Delta E$ (degree of disorder) which is more characteristic for the relaxor ferroelectrics. One can see that $\overline{L}$ and $T_m$ are larger for the ordered films and bulk relaxor, i.e. we have the same conclusions related to the increase of the degree of disorder $\Delta E$ in the sequence ordered films, bulk relaxor ferroelectrics and films of relaxor ferroelectrics.